**On the effect of the path length and transitivity of small-world networks on epidemic dynamics**

Andreas I. Reppas, Konstantinos Spiliotis and Constantinos Siettos[1]

School of Applied Mathematics and Physical Sciences,
National Technical University of Athens, Greece

**Abstract**. We show how one can trace in a systematic way the coarse-grained solutions of individual-based stochastic epidemic models evolving on heterogeneous complex networks with respect to their topological characteristics. In particular, we have developed algorithms that allow the tuning of the transitivity (clustering coefficient) and the average mean-path length allowing the investigation of the "pure" impacts of the two characteristics on the emergent behavior of detailed epidemic models. The framework could be used to shed more light into the influence of weak and strong social ties on epidemic spread within small-world network structures, and ultimately to provide novel systematic computational modeling and exploration of better contagion control strategies.

**1. Introduction**

The dynamic effects of network heterogeneity are central to fundamental public- health measures and policies in order to control the outbreak of a disease in an efficient way through vaccination, quarantine, or use of antiviral drugs on targeted parts of the population.
Hence. the interplay between the emergent dynamics of infectious diseases and the underlying transmission network has been the focus of many studies. In [1], Szendroi and Csányi studied how clustering affects the development of the human immunodeficiency virus epidemics. In [2], Kuperman and Abramson investigated the effect of the rewiring probability used to construct small-world topologies in the dynamics of a simple epidemic model. They showed that above a critical value of this parameter, the dynamics exhibit a phase transition from stationary endemic situations to self-sustained oscillations. In [3], Huang et al. explored the existence of epidemic thresholds in scale-free social networks by modulating the exponent of the distribution. They also studied the influence of the clustering coefficient and mean path length on the epidemic outbreaks. In this, the mean-path length was modulated by changing the size of the network. In [4], Santos et al. investigated the effects of small-world heterogeneities related to the degree distribution. They introduced an algorithm for generating small-world topologies with homogeneous connectivity distributions in order to disassociate small-world properties from connectivity degree heterogeneities. In [5], Shirley and Rushton studied the impact of the structure of four different types of network topologies, namely Erdös–Rèyni, Regular lattices, small-world, and scale-free, on the epidemic dynamics. The authors showed that in scale-free networks exhibiting very small clustering and short path lengths the spread of the disease was faster, while in regular lattices exhibiting high clustering coefficients and long path lengths [6], the spread was the slowest among all other types of networks. Intermediate behavior was observed in small-world topologies characterized by high clustering coefficients and short path-lengths. In [6], Eubank et al., explored the use of dynamic bipartite graphs modeling the real social contact network of Portland, Oregon, USA in order to test targeted vaccination scenarios in the case of a disease outbreak.

In most of these studies, due to the underlying complexity, the analysis is performed by temporal computer simulations: different population sizes along with different network structures-as these are constructed by altering the parameters of the network generation algorithm (such as the rewiring probability in the Watts and Strogatz algorithm [7]) are tested in order to investigate their impact to the emergent dynamics. On one hand, for a more systematic analysis, while one can try to use the tools of statistical physics to write down a coarse-grained master equation to describe the probabilistic time evolution of the macroscopic quantities for simple-structured homogeneous networks, major problems arise in trying to find fair macroscopic models in a closed form when dealing with complex heterogeneous networks. For example, in [8] Pastor-Satorras and Vespignani demonstrated that in the case of scale-free network topologies, building mean-filed approximations based on homogeneity assumptions, may lead to a significant bias on the estimation of the epidemic threshold. Homogeneous network topology characteristics, infinite-size populations are some of the assumptions that may impose both qualitative and quantitative biases at the coarse-grained level. On the other hand, it is has been shown, that real-life structures can significantly deviate from the one generated by standard network-generating algorithms regarding certain topological measurements such as the relation between the clustering and the average path length (see e.g. the discussion in [9, 10]). Furthermore, the topological characteristics of the network may change over time and/or due to the contact duration. For example in [11] the authors reconstructed the contact social network relevant for disease transmission at an American High-School using high-resolution data obtained by wireless sensors. In this, while the clustering coefficient remains almost constant over a wide range of contact durations, the average path length doubles. Hence, there is a need for developing algorithms that are able to tune the topological characteristics in order to match real-world observations (see also [5, 12]). Towards this direction, various algorithms for generating network topologies with prescribed characteristics have been proposed including the tuning of the degree distribution, clustering coefficient and assortativity [13-19].

---

[1] Corresponding author: ksiet@mail.ntua.gr

In this work we focus on the following issue: what are the distinct effects of the average path length (APL) and the transitivity (i.e. the clustering coefficient) (CC) [20] on the emergent dynamics of a simple epidemic model evolving on small-world networks? In order to study the distinct "pure" effects of the APL and CC we developed algorithms for adjusting their values at will, leaving the connectivity distribution and the size of the network invariant. Using these algorithms, one can generate small-world structures possessing, for any practical means, any value for each one of these two parameters. Furthermore, by exploiting Time-Frequency Signal analysis techniques such as Short Time Fourier Transform (STFT) and the Equation-Free approach for multiscale computations [20, 21] we constructed the coarse-grained bifurcation diagrams with respect to the APL and CC. To our knowledge this is the first time that such an analysis is provided. At critical points of the APL (and the CC), while maintaining at the same time all the other topological characteristics at constant values, we found a Hopf-Andronov bifurcation giving rise to self-sustained oscillations a phenomenon that has been also observed in real epidemics [22-24].

**2. Methodology**

Here we show how rewiring coupled with simulated annealing, can be exploited to create network topologies with prescribed values of the CC and the APL. The APL is a global property reflecting the average number of steps required to reach a node from any other node in the network. It is computed taking the mean value of all shortest paths between any two nodes, i.e. $L = \dfrac{\sum d_{i \leftrightarrow j}}{N(N-1)}$, where $d_{i \leftrightarrow j}$ is the shortest path between nodes $i$ and $j$; $N$ is the size of the network. The transitivity expresses a "local" property, that of the formation of cliques and it is calculated by $CC = \dfrac{1}{N}\sum_{i=1}^{N} c_i$, where $c_i = \dfrac{2E_i}{k_i(k_i-1)}$, $k_i$ is the degree of node $i$ and $E_i$ is the number of triangles associated with node $i$.

In order to tune the APL we proceed as follows (see also figure 1):

1) Evaluate the objective function $E(L) = \|L - L^*\|$, where $L^*$ is the target value of the APL.
2) Select randomly two nodes $i$ and $j$ (a) which do not have any common neighbors and (b) each of them has at least one neighbor (say, $i_1$ and $j_1$ respectively) that does not form any triangle with any other neighbor of $i$ and $j$ nor do any common neighbours exist between $i_1$ and $j_1$.
3) Rewire the edges to connect $i$ with $j$ and $i_1$ with $j_1$. At this point the rewiring process may produce multiple components in the network and if it does, the new configuration should be rejected and return back to step 2.
4) Evaluate the new AVP of the network, say $L'$ and the objective function, $E(L')$.
5) Accept or reject the new configuration using the Metropolis procedure.

For the tuning of the CC, two algorithms are used: (a) one which decreases and (b) one which increases its value. For task (a), the procedure can be summarized in the following steps (see also figure 2):

1) Evaluate the objective function $E(C) = \|CC - CC^*\|$ where $C^*$ is the target value of the CC.
2) Select randomly a set of nodes $i$, $j$ with $c_i > 0$ and $c_j > 0$, which do not have any common neighbors.
3) Select the edges that connect the neighbors of the selected nodes, say $i_1, i_2$ for $i$ and $j_1, j_2$ for $j$.
4) Rewire the two edges to connect $i_1$ with $j_1$ and $i_2$ with $j_2$.
5) Evaluate the new clustering coefficient $CC'$ and the objective function $E(CC')$.
6) Accept or reject the new configuration using the Metropolis procedure.

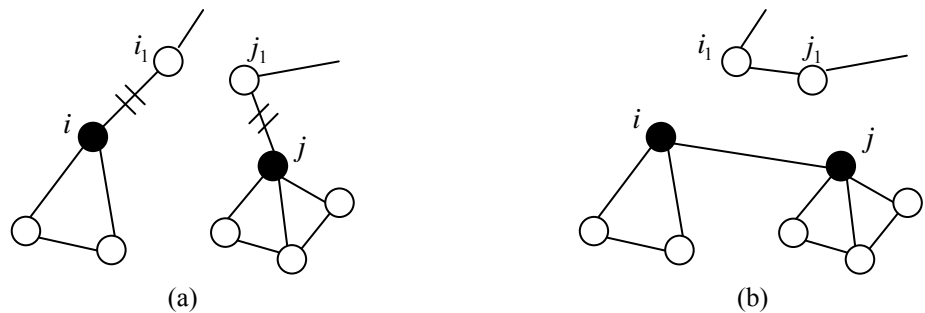

(a)    (b)

Figure 1. The tuning of the path length: (a) step 2 and (b) step 3.

For task (b), steps 2-4 above, are modified as follows:

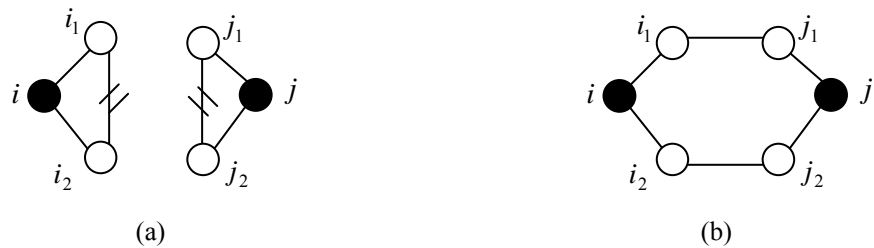

(a)    (b)

Figure 2. The tuning of the CD towards lower values: (a) steps 2, 3 and (b) step 4.

2) Select randomly a node $i$ which has a cycle of length six, i.e. starting visiting the neighbors of $i$ we reach again $i$ in exactly six steps.
3a) Select the node $j$ in the middle of the cycle, i.e. node $j$ should be the third node on the cycle of length six.
3b) Select the neighbors of node $i$, say $(i_1, i_2)$ and node $j$, say $(j_1, j_2)$, which are connected in the current configuration ($i_1$ with $j_1$ and $i_2$ with $j_2$). The neighbors of $i$ are the second and sixth node while the neighbors of $j$ are the third and fifth node on this cycle.
4) Rewire the edges to connect $i_1$ with $i_2$ and $j_1$ with $j_2$. The cycle of length six secures that these nodes are not already connected. At this point the rewiring process may produce multiple components in the network and if it does, the new configuration should be rejected and return back to step 2.

## 3. Simulation Results

For illustration purposes, we use the same epidemiological model appearing in [2]. The epidemic evolves in a discrete time $t$ on small-world networks involving $N$ individuals. The state of each individual is characterized just by its health state respect to the epidemic: (a) Susceptible when the individual is not yet infected but there is a certain probabilistic potential to get infected; (b) Infected when is a carrier of the disease and can potentially transmit it to its links and (c) Recovered when the individual recovers from the infection, cannot transmit and is temporally immunized from the infection.

An infected ndividual $i$ infects a susceptible ink $j$ with a probability $p_{S \to I} = \dfrac{n_i}{k_i}$, where $n_i$ is the number of infected links of the individual $i$. An infected individual recovers in a deterministic way after $\tau_I$ time steps, while a recovered individual becomes susceptible again after $\tau_R$ time steps.

For our simulations we used a total of $N = 10000$ while the initial topologies of the small-world networks were constructed using the Watts and Strogatz algorithm starting from a ring network with $k = 6$ neighbours per node. In [2] it is shown that at a certain value of the rewiring probability $p$ there is a transition from stationary states to synchronization. Implementing the proposed algorithms we were able to investigate the effects of both CC and APL in a distinct manner. In particular, coupling the proposed algorithms with the Equation-Free approach [20,21], using also the STFT we constructed the coarse-grained bifurcation diagrams with respect to APL and CC. Fig. 3a shows the bifurcation diagram of the density of the infected individuals vs. the APL. This was constructed by keeping the CC at a constant value equal to 0.2615. A critical threshold was found at $APL$=7.256. Fig. 3b shows the bifurcation diagram of the density of the infected individuals vs. the CC. This was constructed by keeping the APL at a constant value equal to 6.85. A critical threshold was found at $CC$ = 0.292.

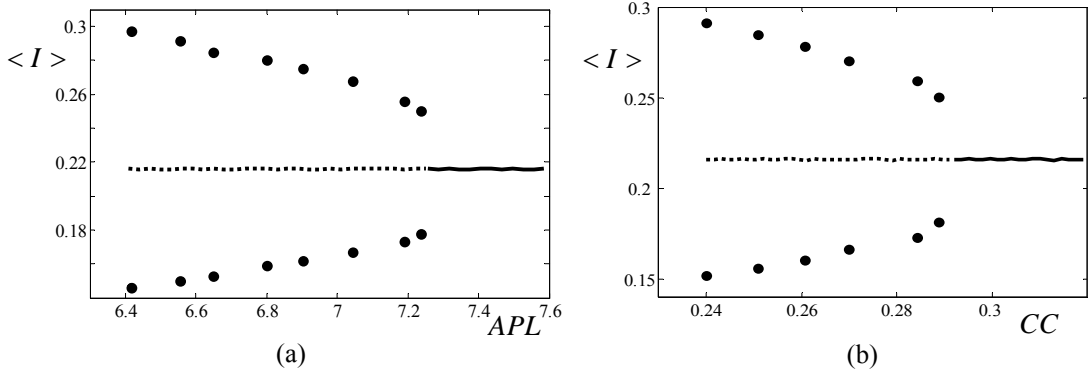

Figure 3. Coarse-grained bifurcation diagram of the density of infected individuals $<I>$ with respect to the (a) APL, with CC=0.2615, (b) CC with APL=6.85; filled circles correspond to the maximum amplitude of oscillations, solid (dotted) lines to stable (unstable) stationary states.

## 4. Conclusions

While the simulation example is admittedly simple, it does demonstrate the scope of the tasks that one can attempt using the proposed framework: it may be used to draw more general conclusions about the "pure" impact of such topological characteristics and based on the detection of topological criticalities that mark the onset of phase transitions, to design more efficient graph-based intervention policies. For example, control measures such as vaccination could be targeted in a distinct way at parts of the population with weak ties (i.e. people with a low transitivity, whose connections govern the value of the average path length), as these are associated with the rapid transmission across the network and synchronization, and at parts of the population with strong ties (i.e. people with a high transitivity) which "inhibit" early spreading [25, 26] but may result to persistent spatial wave-like dynamics [27].